\documentclass[aps,preprint,prd,nofootinbib]{revtex4}
\usepackage{graphicx}
\usepackage{psfrag}

\begin{document}

\title{The Strong Decays of $P-$wave Mixing Heavy-Light $1^+$ States}
\author{Yi Zhang$^{[1]}$, Zhi-Hui Wang$^{[1],[2]}$\footnote{2013086@nun.edu.cn}, Tian-hong Wang$^{[2]}$, Yue Jiang$^{[2]}$,
Qiang Li$^{[3]}$
Guo-Li Wang$^{[2]}$\\}
\address{$^1$School of Electrical $\&$ Information Engineering, North Minzu University, Yinchuan, 750021,\\
$^2$Department of Physics, Harbin Institute of
Technology, Harbin, 150001,\\
$^3$Institute of Theoretical Physics, Chinese Academy of Sciences, Beijing 100190}

 \baselineskip=20pt

\begin{abstract}

 \vspace*{0.5cm}
Many $P-$wave mixing heavy-light $1^+$ states have not been discovered by experiment,
some of them have been discovered but without the information of width, or with large uncertainty widths.
In this paper,
we study the strong decays of $P-$wave mixing heavy-light $1^+$ states $D^0$, $D^\pm$, $D_s^\pm$, $B^0$, $B^\pm$ and $B_s$ by the
improved Bethe-Salpeter(B-S) method in two conditions of mixing angle $\theta$: one is $\theta=35.3^\circ$;
another is considering the correction to mixing angle $\theta=35.3^\circ+\theta_1$.
And we get some valuable predictions of the strong decay widths:
$\Gamma(D_1^{\prime0})=232$ MeV, $\Gamma(D_1^0)=21.5$ MeV,
 $\Gamma(D_1^{\prime\pm})=232$ MeV, $\Gamma(D_1^\pm)=21.5$ MeV,
$\Gamma(D_{s1}^{\prime\pm})=0.0101$ MeV, $\Gamma(D_{s1}^{\pm})=0.950$ MeV,
$\Gamma(B_1^{\prime\pm})=263$ MeV, $\Gamma(B_1^{\pm})=16.8$ MeV,
$\Gamma(B_{s1}^{\prime})=0.01987$ MeV
and $\Gamma(B_{s1})=0.412$ MeV.
We find that the decay widths of $D_{s1}^{\pm}$ and $B_{s1}$ are very sensitive to the mixing angle.
And our results will provide the theoretical
assistance by the future experiments.


\end{abstract}

\maketitle
\section{Introduction}

Heavy-light mesons are very important in the hadronic physics.
During the past several years, a lot of interesting processes
have been obtained for the heavy-light mesons.
Now, many more excited heavy-light mesons have been discovered in experiments.
For the $P$-wave $D_J^*$ mesons,
$D^*_0(2400)^0$, $D_1(2430)^0$, $D_1(2420)^0$, $D^*_2(2460)^0$ and
their charged isospin partners $D^*_0(2400)^\pm$, $D_1(2420)^\pm$, $D^*_2(2460)^\pm$ have been listed in
Particle Data Group(PDG) 2016 edition~\cite{PDG}. Among them, $D_1(2430)^0$ has large
errors in it's decay width and we have few knowledge about it's decay, it's charged isospin partner has not been observed.
Four $P$-wave $D_{sJ}^*$ mesons $D^*_{s0}(2317)^\pm$, $D_{s1}(2460)^\pm$,
$D_{s1}(2536)^\pm$ and $D^*_{s2}(2573)$ were observed in experiments~\cite{PDG}.
The upper bound on the total decay width of the $D^*_{s0}(2317)^\pm$ and $D_{s1}(2460)^\pm$
meson are 3.8 MeV at 95$\%$ confidence level and
3.5 MeV at $95\%$ confidence level~\cite{PDG}, respectively.
The full width of $D_{s1}(2536)^\pm$ is very narrow: $\Gamma=0.92\pm0.05$ MeV.
In 2007, the D0 Collaboration reported two separate excited $B$ mesons
$B_1(5721)^0$ and $B_2^*(5747)^0$ in fully reconstructed decays to $B^{+(*)}\pi^-$~\cite{B1B21}.
The CDF Collaboration also observed two orbitally excited narrow $B^0$ mesons in 2009~\cite{B1B22}.
And they updated the measurement of the properties of orbitally excited $B^0$ and $B_s^0$ mesons in 2015~\cite{update}.
The LHCb Collaboration also gave the precise measurements of the masses and widths of the
$B_1(5721)^{0,+}$ and $B_2^*(5747)^{0,+}$ states in 2015~\cite{B1B2lhcb}.
The CDF Collaboration reported their observations of $B_{s1}(5830)^0$ and $B^*_{s2}(5840)^0$ in 2008~\cite{CDFBS1BS2}.
Later the D0 Collaboration confirmed the existence
of $B^*_{s2}(5840)^0$
and indicated that $B_{s1}(5830)^0$ was not observed with available data~\cite{D0BS1BS2}.
The LHCb Collaboration confirmed the existence of $B_{s1}(5830)^0$ and $B^*_{s2}(5840)^0$ in the $B^{(*)+}K^-$\cite{BS1BS2LHCb}.
The discovery of these excited states
not only enriched the spectroscopy of $P$-wave heavy-light mesons
but also provided us an opportunity to research the properties of $P$-wave heavy-light mesons.

To understand the nature of the $P-$wave heavy-light mesons,
especially the newly observed states, there are a lot of
theoretical efforts to investigate the properties of the $P-$wave heavy-light mesons.
In heavy quark effective theory(HQET)~\cite{hqet1,hqet2},
the angular momentum of light quark $j_q=s_q+L$
($s_q$ and $L$ is the spin and the orbital angular momentum of light quark) is
a good quantum number when the heavy quark has $m_Q\to\infty$ limit, which can
be used to label the states,
so the physical heavy-light states can be described by HQET.
Except HQET, people also studied the mass spectroscopy~\cite{massspectra1,massspectra2,massspectra3,massspectra4} and strong decay of $P-$wave heavy-light mesons by different
methods~\cite{HL1,HL2,HL3,HL4,HL5,HL51,HL6,HL7,HL8,HL9,HL10,HL11,HL12,HL13,HL14}.
The strong decay of $P-$wave heavy-light mesons can helped us to understand the properties
of these mesons and to establish the
heavy-light mesons spectroscopy.

The mesons can be described by the Bethe-Salpeter(B-S) equation.
Ref.~\cite{robert1} took the B-S equation to describe the light mesons $\pi$ and $K$,
then they calculated the mass and decay constant of $\pi$ by the B-S amplitudes~\cite{robert2},
they also studied the weak decays~\cite{robert3} and the strong decays~\cite{robert4} combine the Dyson-Schwinger
equation.
But in this paper, we describe the properties of heavy mesons and
the matrix elements of strong decays by improved B-S method,
which include two improvement~\cite{BS1}: one is about relativistic wavefunctions
which describe bound states with definite quantum number,
and a relativistic form of wavefunctions are solutions of the full Salpeter equations.
The other one is about the matrix elements of strong decays which obtained with
relativistic wavefunctions as input.
So the improved B-S method is good to describe the properties and decays of the heavy mesons with the relativistic corrections.

We have studied the strong decay of $P$-wave $B_s^*$ mesons by
improved B-S method~\cite{Bs}.
We also calculated the productions of $P$-wave mesons in $B,~B_s,~B_c$ mesons~\cite{Pwave1,Pwave2,Pwave3}.
We gave the wavefunctions of mesons by considering
the quantum number $J^P$ or $J^{PC}$ for different states.
$P$-wave $1^+$ states are labelled as $^3P_1$ and $^1P_1$
in our model.
For the unequal
mass system, the $^3P_1$ and $^1P_1$ states are not physical states, the
two physical states $P^{1/2}_1$
and $P_1^{3/2}$,
which are the mixture of $^3P_1$ and $^1P_1$~\cite{hqet1,hqet2}.
In Ref.~\cite{Bs,Pwave1,Pwave2,Pwave3}, we have taken the mixing angle as a definite value $\theta\approx35.3^{\circ}$ for the $P$-wave $1^+$ heavy-light mesons.
However, in fact the heavy quark is not infinitely in $P$-wave $1^+$ states,
considering the correction to the heavy quark limit,
the mixing angle between $^3P_1$ and $^1P_1$ is not a fixed value,
there is a shift which based on the $\theta\approx35.3^{\circ}$,
the shift is different for the different $P$-wave mixing $1^+$ heavy-light states ~\cite{HL1,HL5,mix1}.
In this paper, we will study the strong decays of $P$-wave mixing $1^+$ heavy-light states ($1^{\prime+},~1^+$)
(just like as $D_1^{\prime0}$ and $D^0_1$),
which are in two conditions of mixing angle $\theta$: one is $\theta=35.3^\circ$;
another is considering the correction to mixing angle $\theta=35.3^\circ+\theta_1$,
and talk about the influence of the
shift of the mixing angle on the strong decay of $P$-wave mixing $1^+$ heavy-light states.

The paper is organized as follows.
In Sec.~II, we give the formulation and hadronic matrix element of strong decays;
We show the relativistic wavefunctions of initial mesons and final mesons in Sec.~III;
We talk about the mixing of $^3P_1$ and $^1P_1$ states in Sec.~IV; The corresponding results
and conclusions are present in Sec.~V; Finally in Appendix, we introduce the instantaneous
B-S equation.

\section{The formulation and hadronic matrix element of strong decays}

In this section,
we will show the formulations of the strong decay of $P-$wave mixing states,
and the transition matrix element of the strong decay.
The quantum number of $P-$wave mixing states ($1^{\prime+},~1^+$) both are $1^+$,
considering the kinematic possible mass region,
the ground $P$-wave $1^+$ states only have one strong decay mode: $1^+\to 1^-0^-$.
For the same reason,
we have checked that in the final states of the allowed strong decays
the pseudoscalar $0^-$ state must be the light meson $K,~\pi$,
and the other one $1^-$ state is a heavy meson.

\begin{figure}[htbp]
\centering
\includegraphics[height=5cm]{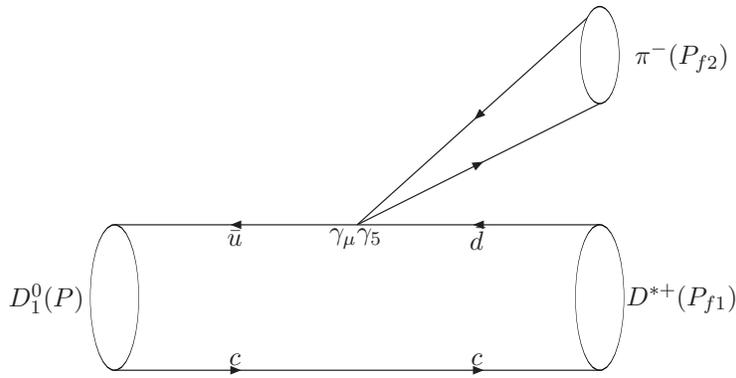}
\caption{\label{strongdecay}The strong decay of $D_1^0\to D^{*+}\pi^-$.}
\end{figure}
\subsection{The strong decay of $1^+$ state}
In order to calculate the strong decay of two mixing states,
taking the channel $D_1^0\to D^{*+}\pi^-$ as an example in Fig.~\ref{strongdecay},
using the reduction formula, PCAC relation and low energy theorem,
the corresponding amplitude can be written as~\cite{Amplitude1,Amplitude2},
\begin{eqnarray}\label{amp}
T(D_1^0\to D^{*+}\pi^-)=\frac{P^\mu_{f2}}{f_\pi}
\langle D^{*+}(P_{f1})|\bar d\gamma_\mu\gamma_5u|D_1^0(P)\rangle\,,
\end{eqnarray}
where $P$, $P_{f1}$, $P_{f2}$ are the momenta of $D_1^0$,
final states $D^{*+}$ and $\pi^-$, respectively.
$f_\pi$ is the decay constant of $\pi^-$.
$\langle D^{*+}(P_{f1})|\bar d\gamma_\mu\gamma_5u|D_1^0(P)\rangle$ is the hadronic matrix element.

With the Eq.~(\ref{amp}), we obtain the strong decay widths formula,
\begin{eqnarray}\label{Gamma}
\Gamma=\frac{|{\vec P}_{f1}|}{24\pi M^2}
\Sigma|T(D_1^0\to D^{*+}\pi^-)|^2
\end{eqnarray}
$M$ is the mass of initial meson $D_1^0$,
${\vec P}_{f1}$ is the three momentum of final heavy meson $D^{*+}$.

Then we find that the most important part in Eq.~(\ref{amp}) is to calculate the hadronic element matrix
$\langle D^{*+}(P_{f1})|\bar d\gamma_\mu\gamma_5u|D_1^0(P)\rangle$.
We will discuss the hadronic element matrix in the next subsection.

\subsection{Hadronic matrix element of strong decays}

In this subsection, we will give the calculation of the hadronic matrix element.
Based on the B-S equation which is a well-known
relativistic method to deal with bound state, we have considered the relativistic
effect, and this new improved method has been developed recently.
Choosing the instantaneous approach, the hadronic
matrix element can be obtained with the Mandelstam formalism~\cite{Mand} and relativistic wave functions.
The numerical values of wavefunctions have been obtained by solving the
full Salpeter equation which we will introduce in Appendix. At the leading order, the hadronic
matrix element can be written as an overlapping integral over the
corresponding initial and final state wavefunctions \cite{BS1},
\begin{eqnarray}\label{a08}
\langle D^{*+}(P_{f1})|\bar d\gamma_\mu\gamma_5u|D_1^0(P)\rangle =
\int\frac{d{\vec{q}_{_f}}}{(2\pi)^3}Tr\left[
\bar{\varphi}^{++}_{_{P_{f1}}}(\vec{q}_{_{f}})\gamma_{\mu}\gamma_5
{\varphi}^{++}_{_{P}}({\vec{q}})\frac{\not\!P}{M}
\right],
\end{eqnarray}
where $P$ and $M$ are the momentum and mass of initial state $D_1^0$,
$\vec q$ and $\vec{q}_{f}=\vec q-\frac{m_c}{m_c+m_d}\vec{P}_{f1}$ are the relative momenta of quark and antiquark
in initial state $D_1^0$ and final state $D^{*+}$, respectively. ${\varphi}^{++}_{_{P}}({\vec{q}})$ and
${\varphi}^{++}_{_{P_{f1}}}(\vec{q}_{_{f}})$ are the positive energy wavefunctions
of $D_1^0$ and $D^{*+}$, which are given in next section.

\section{The Relativistic Wavefunctions}

In our model, improved B-S method, which is based on the constituent
quark model, we give the forms of wave functions by considering
the quantum number $J^P$ or $J^{PC}$ for different states, and
these states in our model are labelled such as: $^3S_1(1^-)$, $^3P_1(1^+)$ and $^1P_1(1^+)$ and so on.
In this paper, we consider the strong decay of
$P-$wave mixing states ($1^{\prime+},~1^+$) which are mixture of $^3P_1(1^+)$ and $^1P_1(1^+)$.
So we only discuss the relativistic wavefunctions of $^1P_1(1^+)$, $^3P_1(1^+)$ and $^3S_1(1^-)$ states.

\subsection{The wavefunctions of $^1P_1$ state}
The general expression for the Salpeter wave function of $^1P_1$ state, which $J^P=1^+$
(or $J^{PC}=1^{+-}$ for quarkonium), can be written as~\cite{Bs,pwave4},

\begin{equation}
\varphi_{1^{+-}}(\vec q)=~ q_\perp\cdot\epsilon\left[f_1(\vec q)+\frac{\not\!{P}}{M}f_2(\vec q)
+\frac{\not\!{q}_\perp}{M}f_3(\vec q)+\frac{\not\!{P}\not\!{q}_\perp}{M^2}f_4(\vec q)\right]\gamma_5
\end{equation}
where $q_\perp=q-\frac{q\cdot P}{M}$ is the relative momentum between quark and anti-quark
in the state, $P$ and $M$ are the momentum and mass of the $1^+$ meson, $\epsilon$ is the polarization vector.
In center of
mass system of the meson, one have $q_\perp = (0, \vec q)$. The wave functions $f_1$, $f_2$, $f_3$ and
$f_4$ which are functions of $q_\perp^2$,
are not independent, they have the relations because
of the constraints equations of Salpeter equation~\cite{Bs,pwave4},
\begin{equation}
f_3=-\frac{M(w_1-w_2)}{m_1w_2+m_2w_1}f_1,
f_4=-\frac{M(w_1+w_2)}{m_1w_2+m_2w_1}f_2
\end{equation}
where $m_1, m_2, w_1, w_2$ are the masses and momentum of the quark and anti-quark in $1^{+-}$ state,
respectively, and $w_1=\sqrt{m_1^2+\vec q^2}$ and $w_2=\sqrt{m_2^2+\vec q^2}$. With this wave function
we can obtain the corresponding positive wave function of $^1P_1$ state,

\begin{eqnarray}\label{1p1}
{\varphi}^{++}_{^1P_1}(\vec q)=q_{\perp}\cdot\epsilon
\left[a_1+a_2\frac{{\not\!P}}{M}+a_3\frac{{\not\!q}_{\perp}}{M}+a_4\frac{{\not\!P}{\not\!q}_{\perp}}{M^2}\right]\gamma_5
\end{eqnarray}
the coefficients $a_1...a_4$ have been defined in Ref. \cite{Bs},

$a_1=\frac{1}{2}(f_1(\vec q)+\frac{w_1+w_2}{m_1+m_2}f_2(\vec q))$, $a_2=\frac{m_1+m_2}{w_1+w_2}a_1$,
$a_3=-\frac{M(w_1-w_2)}{m_1w_2+w_1m_2}a_1$, $a_4=-\frac{M(m_1+m_2)}{m_1w_2+w_1m_2}a_1$.

\subsection{The wavefunctions of $^3P_1$ state}

The general expression for the Salpeter wave function of $^3P_1$ state, which $J^P=1^+$
(or $J^{PC}=1^{++}$ for quarkonium), can be written as~\cite{Bs,pwave4},
\begin{equation}
\varphi_{^3P_1}(\vec q)=i\varepsilon_{\mu\nu\alpha\beta}\frac{P^\nu}{M} q_\perp^\alpha
\epsilon^\beta\left[g_1\gamma^\mu+g_2\frac{\not\!{P}}{M}\gamma^\mu
+g_3\frac{\not\!{q}_\perp}{M}\gamma^\mu+g_4\frac{\not\!{P}\gamma^\mu\not\!{q}_\perp}{M^2}\right]
\end{equation}

According to the relations of the constraints equations of Salpeter equation~\cite{Bs,pwave4},
we have,
\begin{equation}
g_3=-\frac{M(w_1-w_2)}{m_1w_2+m_2w_1}g_1,
g_4=-\frac{M(w_1+w_2)}{m_1w_2+m_2w_1}g_2
\end{equation}

Then, we can get the positive energy wavefunction of $^3P_1$ state~\cite{Bs},

\begin{eqnarray}\label{3p1}
{\varphi}^{++}_{^3P_1}(\vec q)=i\varepsilon_{\mu\nu\alpha\beta}\frac{P^\nu}{M}q^{\alpha}_{\perp}\epsilon^{\beta}\gamma^{\mu}
\left[b_1+b_2\frac{{\not\!P}}{M}+b_3\frac{{\not\!q}_{\perp}}{M}+b_4\frac{{\not\!P}{\not\!q}_{\perp}}{M^2}\right]
\end{eqnarray}
the coefficients $b_1...b_4$ have been defined in Ref. \cite{Bs},

$b_1=\frac{1}{2}(g_1(\vec q)+\frac{w_1+w_2}{m_1+m_2}g_2(\vec q))$, $b_2=-\frac{m_1+m_2}{w_1+w_2}b_1$,
$b_3=\frac{M(w_1-w_2)}{m_1w_2+w_1m_2}b_1$, $b_4=-\frac{M(m_1+m_2)}{m_1w_2+w_1m_2}b_1$.

\subsection{The wavefunctions of $^3S_1$ state}

The general form for the relativistic wavefunction of vector
state $J^P=1^-$(or $J^{PC}=1^{--}$ for quarkonium) can be written
as eight terms, which are constructed by $P_{f1}$, $q_{f\perp}$, $\epsilon_1$ and gamma matrices~\cite{glwang},
\begin{eqnarray}
\varphi_{1^{-}}(\vec q_f)&=&
q_{f\perp}\cdot{\epsilon}_1
\left[f'_1+\frac{\not\!P_{f1}}{M_{f1}}f'_2+
\frac{{\not\!q}_{f\perp}}{M_{f1}}f'_3+\frac{{\not\!P_{f1}}
{\not\!q}_{f\perp}}{M_{f1}^2} f'_4\right]+M_{f1}{\not\!\epsilon_1}f'_5\\ \nonumber
&+&
{\not\!\epsilon_1}{\not\!P_{f1}}f'_6+
({\not\!q}_{f\perp}{\not\!\epsilon_1}-
q_{f\perp}\cdot{\epsilon_1}
f'_7+\frac{1}{M_{f1}}({\not\!P_{f1}}{\not\!\epsilon_1}
{\not\!q}_{f\perp}-{\not\!P_{f1}}q_{f\perp}\cdot{\epsilon_1})
f'_8,\label{eq13}
\end{eqnarray}
where ${\epsilon}_1$ is the polarization vector of the
vector meson in the final state.

According to the relations of the constraints equations of Salpeter equation~\cite{Bs,pwave4},
we have,
$$f'_1=\frac{\left[q_{f\perp}^2 f'_3+M_{f1}^2f'_5
\right] (m'_1m'_2-w'_1w'_2+q_{f\perp}^2)}
{M_{f1}(m'_1+m'_2)q_{f\perp}^2},~~~f'_7=\frac{f'_5(q_{f\perp})M_{f1}(-w'_1+w'_2)}
{(m'_1w'_2+m'_2w'_1)},$$
$$f'_2=\frac{\left[-q_{f\perp}^2 f'_4+M_{f1}^2f'_6(q_{f\perp})\right]
(m'_1w'_2-m'_2w'_1)}
{M_{f1}(w'_1+w'_2)q_{f\perp}^2},~~~f'_8=\frac{f'_6M_{f1}(w'_1w'_2-m'_1m'_2-q_{f\perp}^2)}
{(m'_1+m'_2)q_{f\perp}^2}.$$

The relativistic positive wavefunction of $^3S_1$ state can be written as:
\begin{eqnarray}
{\varphi}_{1^{-}}^{++}(\vec{q_f})&=&b_1\not\!{\epsilon}_1+b_2\not\!{\epsilon}_1\not\!{P_{f1}}
+b_3(\not\!{q_{f\bot}}\not\!{\epsilon}_1-q_{f\bot}\cdot{\epsilon}_1)
+b_4(\not\!{P_{f1}}\not\!{\epsilon}_1\not\!{q_{f\bot}}-\not\!{P_{f1}}q_{f\bot}\cdot{\epsilon}_1)
\nonumber\\
&&+q_{f\bot}\cdot{\epsilon}_1(b_5+b_6\not\!{P_{f1}}+b_7\not\!{q_{f\bot}}+b_8\not\!{q_{f\bot}}\not\!{P_{f1}}),
\end{eqnarray}
where we first define the parameter $n_i$ which are the functions of
$f'_i$ ($^3S_1$ wave functions):
$$n_1=f'_5
-f'_6\frac{(w'_1+w'_2)}{(m'_1+m'_2)}, n_2=f'_5
-f'_6\frac{(m'_1+m'_2)}{(w'_1+w'_2)}, n_3=f'_3
+f'_4\frac{(m'_1+m'_2)}{(w'_1+w'_2)},$$ then we define
the parameters $b_i$ which are the functions of $f'_i$ and $n_i$:
$$b_1=\frac{M_{f1}}{2}n_1, b_2=-\frac{(m'_1+m'_2)}{2(w'_1+w'_2)}n_1,
 b_3=\frac{M_{f1}(w'_2-w'_1)}{2(m'_1w'_2+m'_2w'_1)}n_1, b_4=\frac{(w'_1+w'_2)}{2(w'_1w'_2+m'_1m'_2-{q_{f\bot}^{2}})}n_1,$$
 $$b_5=\frac{1}{2M_{f1}}\frac{(m'_1+m'_2)(M_{f1}^2n_2+{q_{f\bot}^{2}}n_3)}{(w'_1w'_2+m'_1m'_2+{q_{f\bot}^{2}})},
  b_6=\frac{1}{2M_{f1}^2}\frac{(w'_1-w'_2)(M_{f1}^2n_2+{q_{f\bot}^{2}}n_3)}{(w'_1w'_2+m'_1m'_2+{q_{f\bot}^{2}})},$$
$$b_7=\frac{n_3}{2M_{f1}}-\frac{f'_6M_{f1}}{(m'_1w'_2+m'_2w'_1)},
 b_8=\frac{1}{2M_{f1}^2}\frac{w'_1+w'_2}{m'_1+m'_2}n_3-f'_5\frac{w'_1+w'_2}{(m'_1+m'_2)(w'_1w'_2+m'_1m'_2-{q_{f\bot}^{2}})}.$$

\section{The Mixing of $^3P_1$ and $^1P_1$ states}

In a heavy-light bound states, there are two conserved quantities, one is the the
spin $S_Q$ of the heavy quark, another is the total angular momentum $j_q=s_q+L$ of
the light quark, and the whole angular momentum $J$ of the corresponding meson
is sum of $S_Q$ and $j_q$. So there are two physical $P$-wave heavy-light mixing states $P^{1/2}_1$
and $P_1^{3/2}$.

In our model, we give expressions of the wavefunctions in term of the quantum
number $J^P$ (or $J^{PC}$) which are very good to
describe the equal mass systems in heavy mesons.
There are two physical $P$-wave states $^3P_1(1^{++})$ and $^1P_1(1^{+-})$ for the equal mass systems,
but they are not the physical states when there is no
the charge conjugation parity for unequal mass system.
In the heavy quark limit,
the physical states $P^{1/2}_1$ and $P_1^{3/2}$ of $P$-wave heavy-light mesons can be written as\cite{hqet1,hqet2,mix-rosner},
\begin{eqnarray}\label{mix-old}
&&|P^{3/2}_1>=\sqrt{\frac{2}{3}}|^1P_1>+\sqrt{\frac{1}{3}}|^3P_1>\\ \nonumber
&&|P^{1/2}_1>=-\sqrt{\frac{1}{3}}|^1P_1>+\sqrt{\frac{2}{3}}|^3P_1>,
\end{eqnarray}
where the Eq.~(\ref{mix-old}) is the same as the result of Ref.~\cite{Bs,Pwave3} if we take the $\theta=35.3^\circ$.
So the wavefunctions for physical $P^{1/2}_1$ and $P_1^{3/2}$
states can be obtained by these mixing
relations of the $^3P_1$ and $^1P_1$ wavefunctions which are shown in Section.~III.
However, when we solve the B-S equation, the mass of heavy quark is a fixed value which is not infinitely,
consider the correction to the heavy quark limit,
the physical states as $1^{\prime+}$ and $1^+$ which are the mixture of $P^{1/2}_1$ and $P_1^{3/2}$~\cite{HL1,HL5,mix1},
\begin{eqnarray}\label{mix-new}
&&|1^+>=\cos\theta_1|P^{3/2}_1>+\sin\theta_1|P^{1/2}_1>\\ \nonumber
&&|1^{\prime+}>=-\sin\theta_1|P^{3/2}_1>+\cos\theta_1|P^{1/2}_1>,
\end{eqnarray}
when in  the heavy quark limit, e.g: $\theta_1=0^\circ$,
$|1^{\prime+}>=|P^{1/2}_1>$ and $|1^{+}>=|P^{3/2}_1>$.
For the $1^+$ states $D$ and $D_s$, $\theta_1=-(0.1\pm0.05)$ rad=$-(5.7\pm2.9)^\circ$~\cite{HL1,HL5},
for the $1^+$ states $B$ and $B_s$, $\theta_1=-(0.03\pm0.015)$ rad=$-(1.72\pm0.86)^\circ$~\cite{HL5}.
Taking the Eq.~(\ref{mix-old}) into Eq.~(\ref{mix-new}),
we can get the relation of $1^{\prime+}$,$1^+$ and $^3P_1$, $^1P_1$,
\begin{eqnarray}\label{mix-new-1}
&&|1^+>=\cos\theta|^1P_1>+\sin\theta|^3P_1>\\ \nonumber
&&|1^{\prime+}>=-\sin\theta|^1P_1>+\cos\theta|^3P_1>,
\end{eqnarray}
where $\theta=35.3^\circ+\theta_1$.
According to the Eq.~(\ref{mix-new-1}),
we have obtained the mass spectra of the $P$-wave mixing states $1^{\prime+}$ and $1^+$ by
solving the full Salpeter equation in Table.~\ref{massspectra}.

\begin{table}[htbp]
\caption{ \label{massspectra}The mass spectra of the $P$-wave mixing states $1^{\prime+}$ and $1^+$ in the units of MeV.
`Ex.' means the experimental data from
PDG~\cite{PDG}, and `Th.' means our prediction.}
\begin{center}
\begin{tabular}{ccccccccc}
\hline \hline
States&Th.&Ex.~&States&Th.&Ex.&States&Th.&Ex.  \\
\hline
$D^{\prime0}_1$&2427.0&$2427\pm26\pm25$&$D^{\prime\pm}_{s1}$&2460.0&$2459.6\pm0.9$&$B_1^{\prime0}$&5710.0&--\\
$D_1^0$&2422.0&$2421.4\pm0.6$&$D^\pm_{s1}$&2536.0&$2535.18\pm0.24$&$B^0_1$&5726.0&$5726.0\pm1.3$\\
\hline
$D^{\prime\pm}_1$&2427.0&--&$B^\prime_{s1}$&5820.0&--&$B_1^{\prime\pm}$&5710.0&--\\
$D_1^\pm$&2422.0&$2423.2\pm2.4$&$B_{s1}$&5829.0&$5828.78\pm0.35$&$B_1^{\pm}$&5726.0&$5726.8^{+3.2}_{-4.0}$\\
\hline \hline
\end{tabular}
\end{center}
\end{table}

\section{Number results and discussions}

In order to fix Cornell potential in Eq.(\ref{eq16}) and masses of quarks,
 we take these parameters: $a=e=2.7183,
\lambda=0.21$ GeV$^2$, ${\Lambda}_{QCD}=0.27$ GeV, $\alpha=0.06$
GeV, $m_u=0.305$ GeV, $m_d$=0.311 GeV, $m_s$=0.500 GeV, $m_b=4.96$ GeV, $m_c=1.62$ GeV, $etc$~\cite{mass1},
which are best to fit the mass spectra of ground states $B$, $D$ mesons and other heavy mesons.
And we get the masses of ground states: $M_{D^{*0}}=2.007$ GeV,
$M_{D^{*\pm}}=2.010$ GeV, $M_{D_s^{*\pm}}=2.112$ GeV, $M_{B^*}=5.325$ GeV,
$M_{B^*_s}=5.415$ GeV.
For the light mesons, the masses and decay constants are:
$M_\pi=0.140$ GeV,      $f_\pi=0.130$ GeV,
$M_K=0.494$ GeV,      $f_K=0.156$ GeV~\cite{PDG}, respectively.

\subsection{$D_1^{\prime0}$, $D^0_1$ and $D_1^{\prime\pm}$, $D^{\pm}_1$}

\begin{table}[htbp]
\caption{\label{D0D10}The decay widths of two-body strong decays of $D^0$ mixing states $1^{\prime+}$ and $1^+$ in the units of MeV and
$\theta_1=-(0.1\pm0.05)$ rad=$-(5.7\pm2.9)^\circ$~\cite{HL1,HL5}.}
\begin{center}
\begin{tabular}{|c|c|c|c|c|c|c|c|}
\hline \hline
Mode&$\theta=35.3^{\circ}$&$\theta=35.3^{\circ}+\theta_1$&\cite{HL3}&\cite{HL6}&\cite{HL7}&\cite{HL9}&\cite{PDG}   \\
\hline
$D^{\prime0}_1\to D^*\pi$&232&228$\sim$232&--&244&272&220&$384^{+107}_{-75}\pm74$\\
\hline
$D_1^0\to D^*\pi$&17.3&17.6$\sim$21.5&11&25&22&21.6&$27.4\pm2.5$\\
\hline \hline
\end{tabular}
\end{center}
\end{table}

\begin{figure}[htbp]
\centering
\includegraphics[height=6cm]{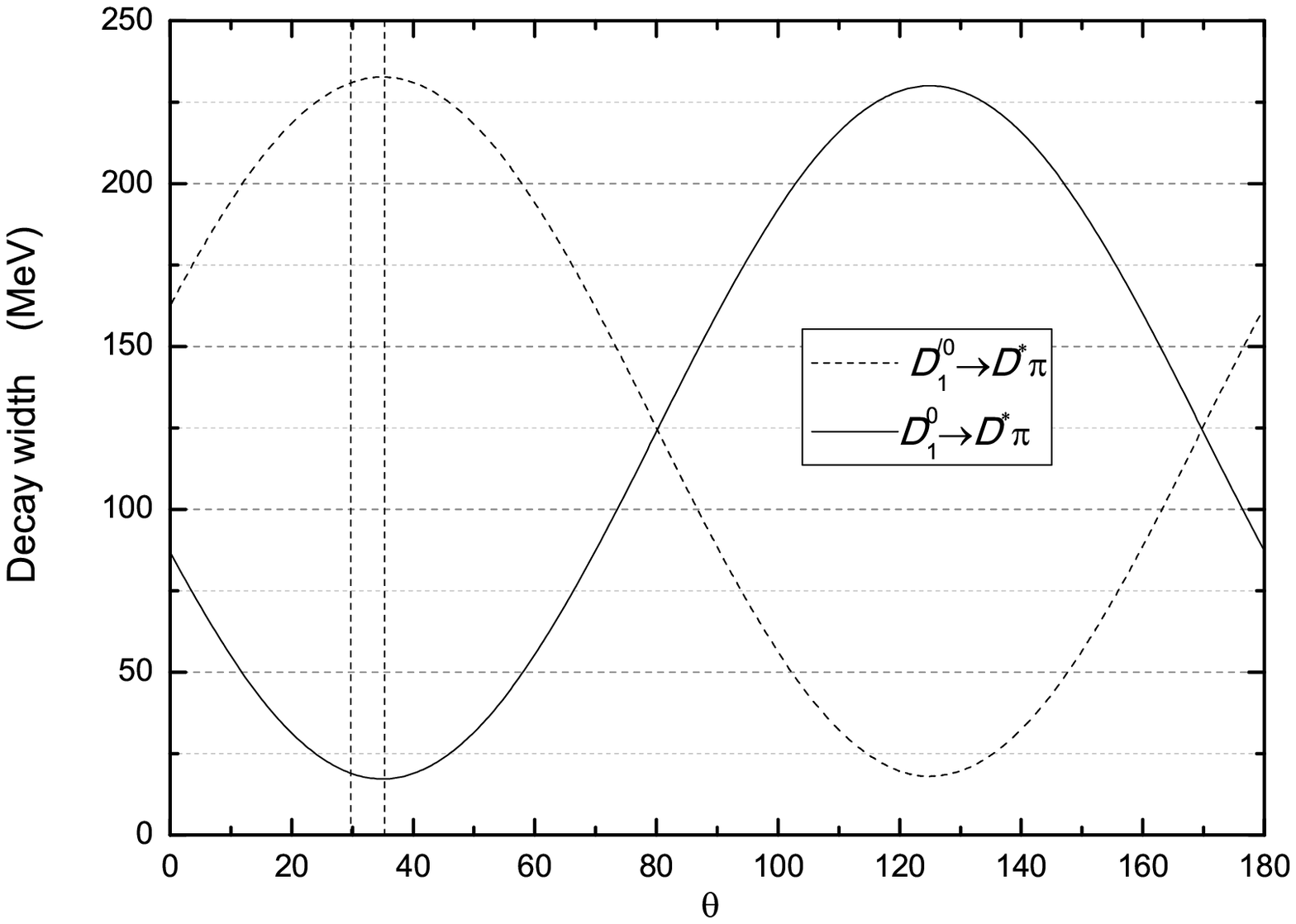}
\caption{\label{D1D11-Theta}The decay widths of $D_1^{\prime0}$ and $D_1^0$.}
\end{figure}

\begin{table}[htbp]
\caption{\label{D1D11}The decay widths of two-body strong decays of $D^{\pm}$ mixing states $1^{\prime+}$ and $1^+$ in the units of MeV and
$\theta_1=-(0.1\pm0.05)$ rad=$-(5.7\pm2.9)^\circ$~\cite{HL1,HL5}.}
\begin{center}
\begin{tabular}{|c|c|c|c|c|c|c|c|}
\hline \hline
Mode&$\theta=35.3^{\circ}$&$\theta=35.3^{\circ}+\theta_1$&\cite{HL3}&\cite{HL6}&\cite{HL7}&\cite{HL9}&\cite{PDG}   \\
\hline
$D^{\prime\pm}_1\to D^*\pi$&232&228$\sim$232&--&244&272&220&--\\
\hline
$D_1^{\pm}\to D^*\pi$&17.2&18.1$\sim$22.0&11&25&22&21.6&$25\pm6$\\
\hline \hline
\end{tabular}
\end{center}
\end{table}

The $D_1^{\prime0}$, $D_1^0$ and $D_1^{\pm}$ have been listed in PDG~\cite{PDG},
$D_1^{\prime0}$ is a broad state with large uncertainty,
$D_1^{0}$ and $D_1^{\pm}$ are narrow states with small uncertainty.
But there is no evidence of another state $D_1^{\prime\pm}$ in experiment.
So we predict the mass of $D_1^{\prime\pm}$ which is the same as $D_1(2430)^0$ by the improved the B-S method in Table.~\ref{massspectra}.
And the two body strong decays of these states only happened in $D^*\pi$ channel which is OZI allowed.

We calculate the transition matrix elements by
the wavefunctions numerically, and get the strong decay widths of $D_1^{\prime0}$ and $D^0_1$
which are predicted by us and other authors in Table.~\ref{D0D10}.
And we show the results in two condition of the mixing angle $\theta$:
one is $\theta=35.3^\circ$,
another which consider the correction $\theta=35.3^\circ+\theta_1$.
We find that the results of $D^{\prime0}_1\to D^*\pi$ are very close in two conditions,
and both of them are smaller than the center of the experiment value,
but if we consider the large uncertainty of the experiment value for $D^{\prime0}_1$,
our results are reasonable.
We also find that our results are consistent with the results of Ref.~\cite{HL6} and Ref.~\cite{HL9},
but smaller than the result of Ref.~\cite{HL7}.
Though the predicted masses of the $P$-wave mixing states are
similar for different models, the predicted decay widths are much
different.
The situation is similar in other channels. For example,
our prediction of $D_1^{0}\to D^*\pi$, when we consider the correction to the heavy quark limit,
the decay width is increased which is consistent with the results of other model and close to the lower limit of experimental value.
In Fig.~\ref{D1D11-Theta}, we plot the relation of mixing angle $\theta$ to the strong decay width of $D_1^{\prime0}$ and $D_1^0$.
The $D_1^{\prime0}$ meson is broad state and the influence of mixing angle is very small.
The width of $D_1^{0}$ meson is at bottom of curve which is sensitive to mixing
angle.
We determine the mixing angle $\theta=35.3^\circ+\theta_{1min}=26.7^\circ$ which is the best description of experimental value.
But for $D_1^{\prime0}$ meson, there are large uncertainties with experimental value,
many more experimental confirmations is needed in the future.
In contrast, we also give the strong decay widths of $D_1^{\prime\pm}$ and $D^\pm_1$ in Table.~\ref{D1D11}.
The strong decay widths of $D^\pm_1$ and $D_1^{\prime\pm}$ are very similar to the results of $D_1^{\prime0}$ and $D_1^0$,
because of the masses of light quark in $P$-wave mixing states are very close: $m_u\approx m_d$.
The strong decay of $D^{\prime\pm}_1$ also provide a good way to observe this meson in experiment.

\subsection{$D^{\prime\pm}_{s1}$ and $D^\pm_{s1}$}

\begin{table}[htbp]
\caption{\label{DsDs1}The decay widths of two-body strong decays of $D_s$ mixing states $1^{\prime+}$ and $1^+$ in the units of MeV and
$\theta_1=-(0.1\pm0.05)$ rad=$-(5.7\pm2.9)^\circ$~\cite{HL1,HL5}.}
\begin{center}
\vspace{0.2em}\centering\footnotesize
\begin{tabular}{|c|c|c|c|c|c|c|c|c|c|}
\hline \hline
Mode&$\theta=35.3^{\circ}$&$\theta=35.3^{\circ}+\theta_1$&\cite{HL2}&\cite{HL3}&\cite{HL6}&\cite{HL7}&\cite{HL9}&\cite{HL11}&\cite{PDG}\\
\hline
$D^{\prime\pm}_{s1}\to D_s^*\pi$&0.0101&0.00984$\sim$0.0100&0.0215&--&$\sim 0.010$&--&--&0.01141&$<3.5$\\
\hline
$D^\pm_{s1}\to D^*K$&0.449&0.950$\sim$5.46&--&$<1$&0.340&0.800&0.350&--&$0.92\pm0.05$\\
\hline \hline
\end{tabular}
\end{center}
\end{table}

\begin{figure}[htbp]
\centering
\includegraphics[height=5cm]{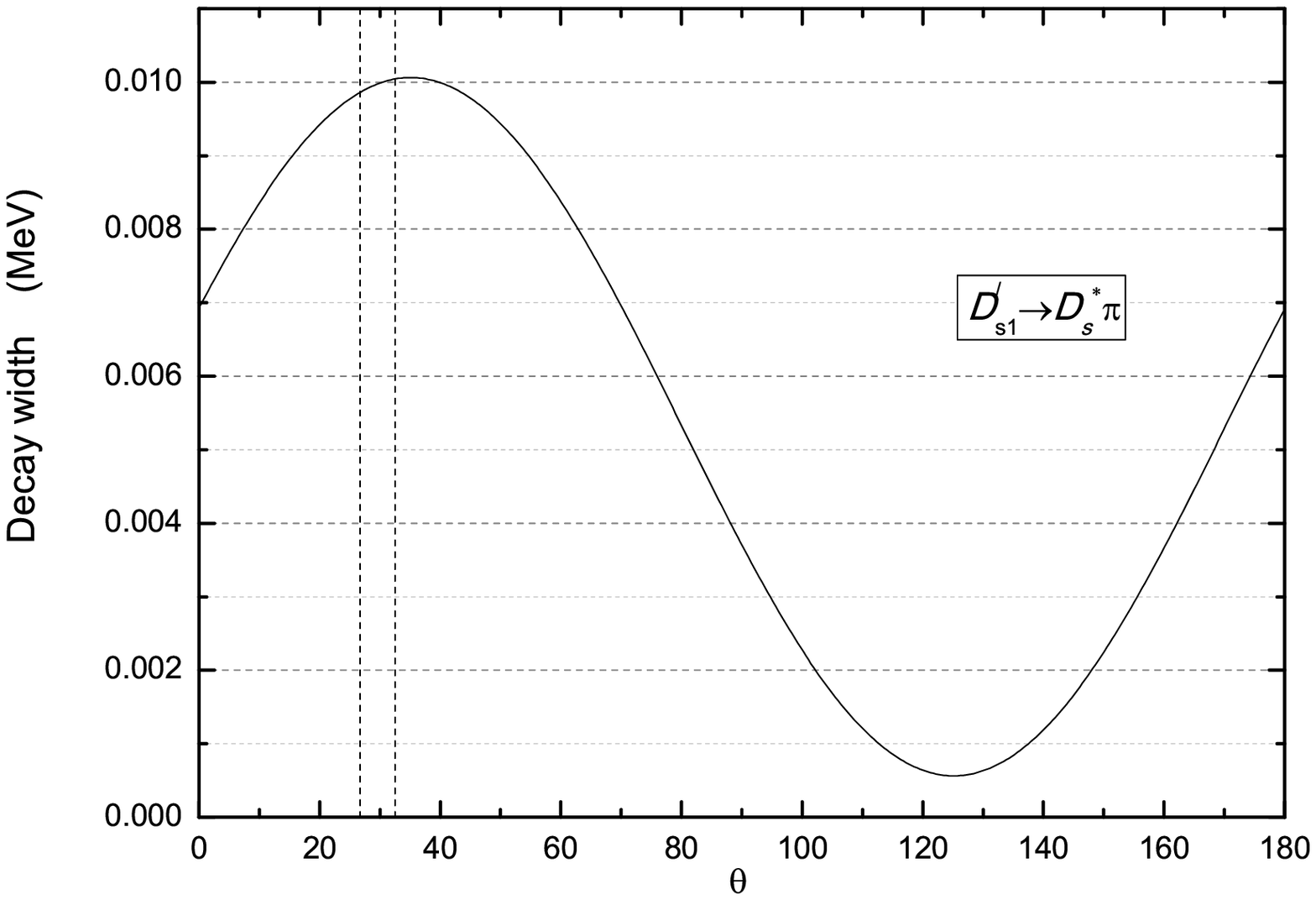}
\includegraphics[height=5cm]{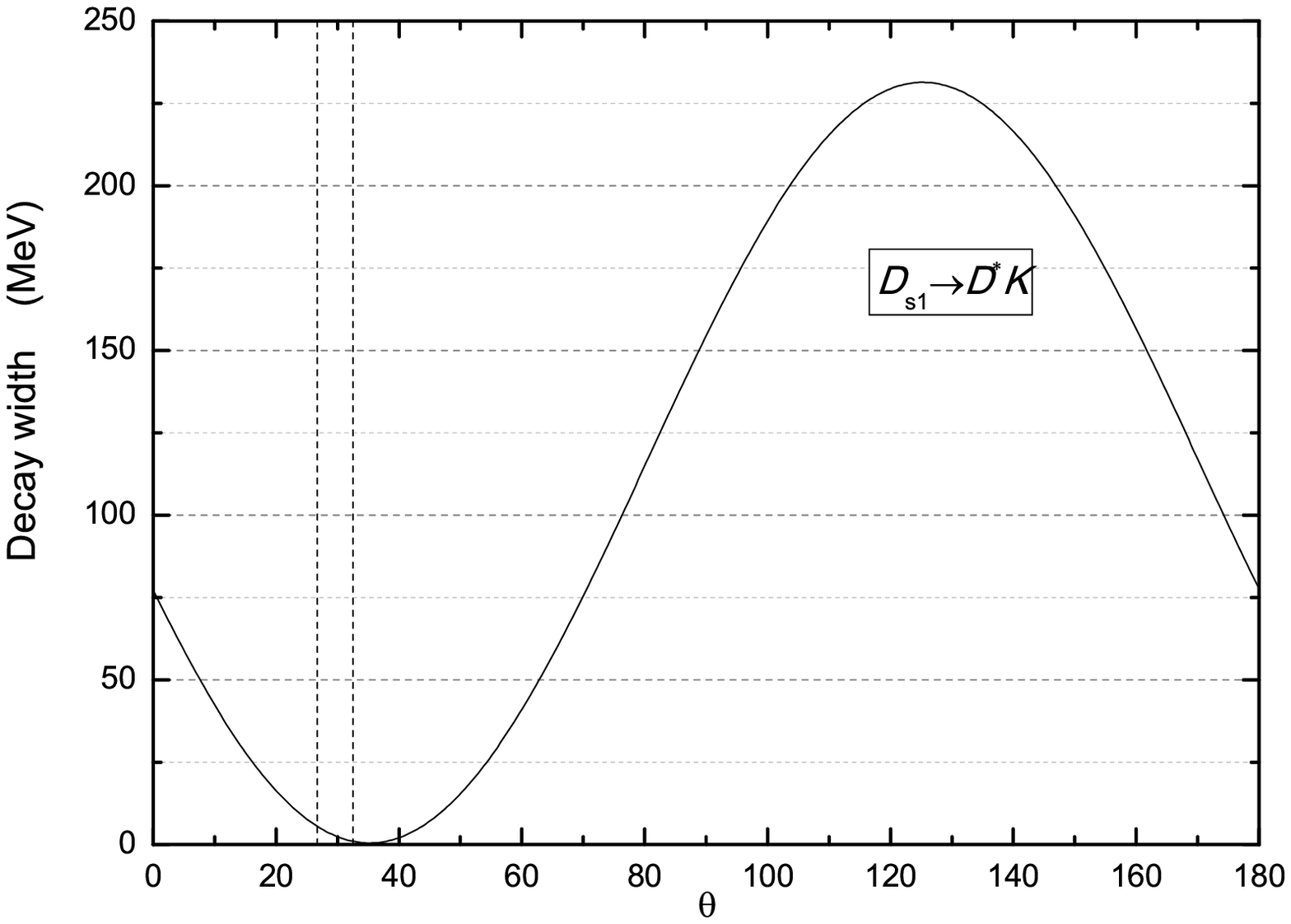}
\caption{\label{Ds1Ds11-Theta}The decay widths of $D^{\prime\pm}_{s1}$ and $D^\pm_{s1}$.}
\end{figure}

Both $D^{\prime\pm}_{s1}$ and $D^\pm_{s1}$ have small width.
The $D^{\prime\pm}_{s1}$ meson is below the threshold of $D^*$ and $K$,
so the OZI allowed strong decay is forbidden,
and the dominant strong decay channel of $D^{\prime\pm}_{s1}$ meson is the isospin
symmetry violating decay via $\pi^0-\eta$ mixing which is $D^{\prime\pm}_{s1}\to D_s^{*\pm}\eta\to D_s^{*\pm}\pi^0$~\cite{pieta}.
Because of the small value of mixing parameter $t_{\pi\eta}=<\pi^0|\mathcal H|\eta>=-0.003$ GeV$^2$~\cite{pieta},
the decay is suppressed much heavily, and the decay widths is very narrow.
The mass of $D^{\pm}_{s1}$ is larger than the threshold of $D^*K$,
but the width is also very narrow because of the small kinematic range.

We have given two results of our method in Table.~\ref{DsDs1}:
(1) $\theta=35.3^\circ$;
(2) considering the correction $\theta=35.3^\circ+\theta_1$.
For the strong decay $D^{\prime\pm}_{s1}\to D_s^*\pi$,
the results are very similar in two conditions and close to the results of Ref.~\cite{HL6} and Ref.~\cite{HL11}.
The results of $D^{\prime\pm}_{s1}\to D_s^*\pi$ are smaller than the result of Ref.~\cite{HL2},
but they are reliable compare with experimental result.
We find that the results of $D^{\pm}_{s1}\to D^*K$ are sensitive to the mixing angle $\theta$.
When the mixing angle $\theta=35.3^\circ$,  the decay width of $D^{\pm}_{s1}$ is close to the results of Ref.~\cite{HL6} and Ref.~\cite{HL9},
but smaller than the result of Ref.~\cite{HL7} and the experimental value in Ref.~\cite{PDG}.
With the correction to mixing angle,
we get the decay width of $D^{\pm}_{s1}$ as: $\Gamma(D^{\pm}_{s1})=0.950\sim5.46$ MeV.
In order to compare with the experimental data,
we plot the relation of strong decay width $D^{\prime\pm}_{s1}$ and $D^\pm_{s1}$ vs the mixing angle $\theta$ in Fig.~\ref{Ds1Ds11-Theta}.
We find that the results of $D^{\pm}_{s1}$ are at bottom of curve and close to zero with both of two conditions,
so the result of $D^{\pm}_{s1}$ is sensitive to mixing angle.
It shows that with $\theta=32.5^\circ$,
the corresponding result $\Gamma(D^{\pm}_{s1})=0.950$ MeV,
which is consistent with the experimental data of $D_{s1}(2536)^{\pm}$: $\Gamma=0.92\pm0.05$ MeV~\cite{PDG}.

\subsection{$B_1^{\prime0}$, $B_1^0$ and $B_1^{\prime\pm}$, $B_1^{\pm}$}

\begin{table}[htbp]
\caption{\label{B0B10}The decay widths of two-body strong decays of $B^0$ mixing states $1^{\prime+}$ and $1^+$  in the units of MeV
and $\theta_1=-(0.03\pm0.015)$ rad=$-(1.72\pm0.86)^\circ$~\cite{HL5}.}
\begin{center}
\vspace{0.2em}\centering
\begin{tabular}{|c|c|c|c|c|c|c|}
\hline \hline
Mode&$\theta=35.3^{\circ}$&$\theta=35.3^{\circ}+\theta_1$&\cite{HL51}&\cite{HL9}&\cite{HL12}&\cite{PDG}\\
 \hline
$B_1^{\prime0}\to B^*\pi$&262.4&262.1$\sim$262.4&250&219&139&--\\
\hline
$B_1^0\to B^*\pi$&15.6&15.6$\sim$16.0&--&30&20&23$\pm3\pm4$\\
\hline \hline
\end{tabular}
\end{center}
\end{table}

\begin{table}[htbp]
\caption{\label{B1B11}The decay widths of two-body strong decays of $B^{\pm}$ mixing states $1^{\prime+}$ and $1^+$  in the units of MeV
and $\theta_1=-(0.03\pm0.015)$ rad=$-(1.72\pm0.86)^\circ$~\cite{HL5}.}
\begin{center}
\vspace{0.2em}\centering\small
\begin{tabular}{|c|c|c|c|c|c|c|}
\hline \hline
Mode&$\theta=35.3^{\circ}$&$\theta=35.3^{\circ}+\theta_1$&\cite{HL51}&\cite{HL9}&\cite{HL12}&\cite{PDG}\\
 \hline
$B_1^{\prime\pm}\to B^*\pi$&263&262.4$\sim$262.9&250&219&139&--\\
\hline
$B_1^{\pm}\to B^*\pi$&16.0&16.1$\sim$16.8&--&30&20&$31\pm6$\\
\hline \hline
\end{tabular}
\end{center}
\end{table}

\begin{figure}[htbp]
\centering
\includegraphics[height=6cm]{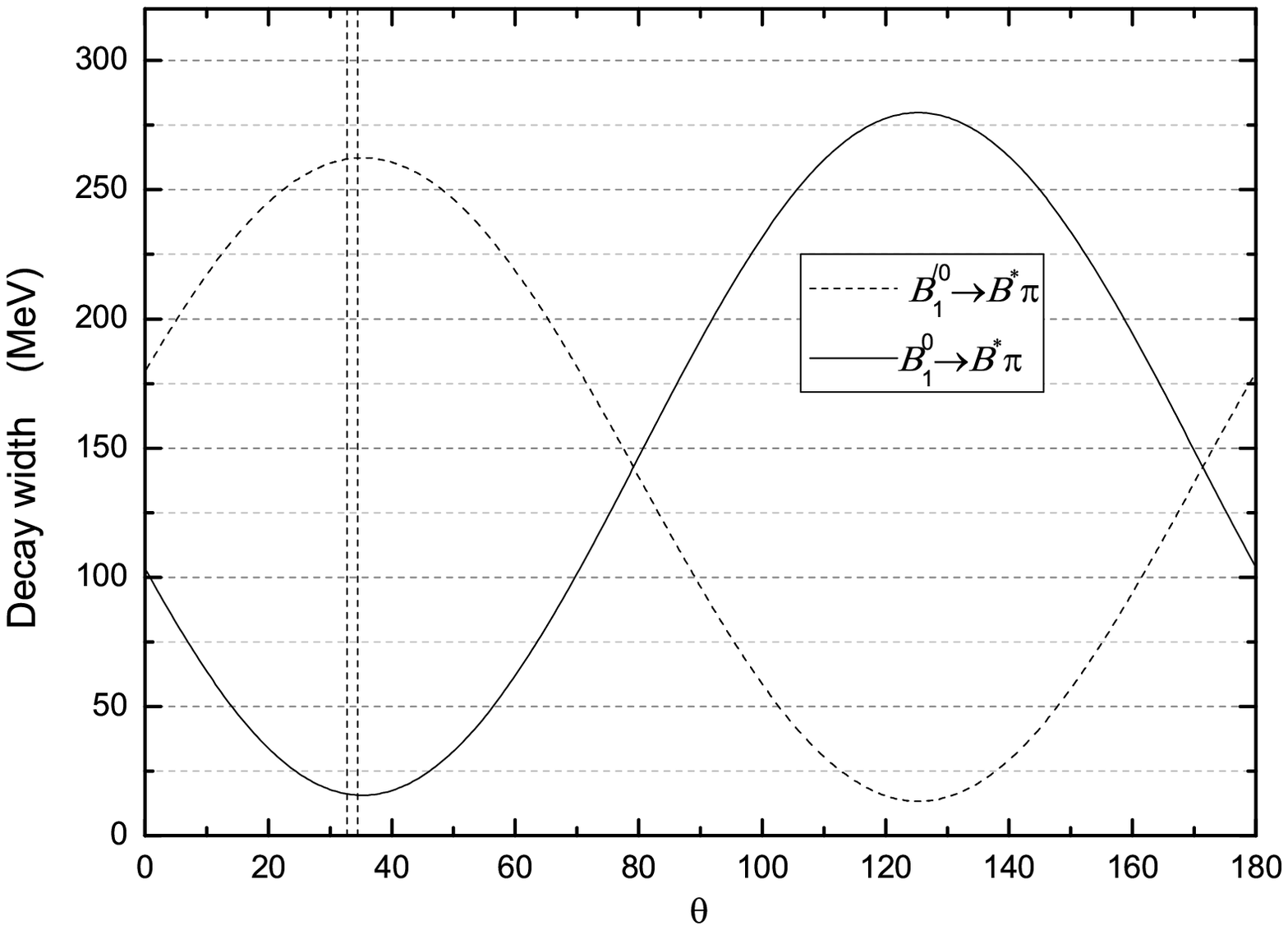}
\caption{\label{B1B11-Theta}The decay widths of $B_1^{\prime0}$ and $B_1^0$.}
\end{figure}

Experiment only observed $B_1^{0}$ and  $B_1^{\pm}$ which were considered
as $B_1(5721)^0$ and $B_1(5721)^+$~\cite{B1B21,B1B22,update},
and there are no evidence for the $B_1^{\prime0}$ and $B_1^{\prime\pm}$.
So we predict the masses of $B_1^{\prime0}$ and $B_1^{\prime\pm}$ firstly,
then calculate the strong decay of these mesons.
Because of large kinematic range,
all of these four states have the OZI allowed strong decay channels.

We show the strong decay widths for the $P$-wave mixing $B$ mesons in two conditions which are the same as $P$-wave mixing $D$ mesons
in Table.~\ref{B0B10} and Table.~\ref{B1B11}.
For comparison, we list the experimental data and some results of other model predictions and plot
the relations of the decay widths of $B_1^{\prime0}$ and $B_1^0$ vs the mixing angle $\theta$ in Fig.~\ref{B1B11-Theta}.
For $B_1^{\prime0}$ and $B_1^{\prime\pm}$, the decay widths of two conditions are very close,
which are consistent with the rsult of QCD sum rules: $\Gamma(B^{\prime}_1)\simeq250$ MeV~\cite{HL51}.
The decay widths of $B_1^0$ and $B_1^\pm$ are  close to the result of Ref.~\cite{HL51}
and smaller than the results of Ref.~\cite{HL9} and Ref.~\cite{HL12},
the results of  $B_1^{\pm}$ is also much smaller than the experimental value,
but the results of  $B_1^{0}$ are very close to the upper limits of experimental value.
Because the experimental results of $B_1^{\pm}$ have large uncertainties,
there will need to be confirmed in future in experimentally,
and our result will provide the theoretical assistance.

\subsection{$B_{s1}^\prime$ and $B_{s1}$}

\begin{table}[htbp]
\caption{\label{Bs1Bs11}The decay widths of two-body strong decays of $B_s$ mixing states $1^{\prime+}$ and $1^+$ in the units of MeV
and $\theta_1=-(0.03\pm0.015)$ rad=$-(1.72\pm0.86)^\circ$~\cite{HL5}.}
\begin{center}
\vspace{0.2em}\centering\scriptsize
\begin{tabular}{|c|c|c|c|c|c|c|c|c|c|}
\hline \hline
Mode&$\theta=35.3^{\circ}$&$\theta=35.3^{\circ}+\theta_{1}$&\cite{HL2}&\cite{HL3}&\cite{HL9}&\cite{HL11}&\cite{HL13}&\cite{HL14}&\cite{PDG}\\
\hline
$B^\prime_{s1}\to B_s^*\pi$&0.01987&0.01986$\sim$0.01987&0.0215&--&--&0.01036&--&--&-- \\
\hline
$B_{s1}\to B^*K$&0.0396&0.0834$\sim$0.412&--&$<1$&$0.4\sim1$&--&$0.7\pm0.3\pm0.3$&0.098&$0.5\pm0.3\pm0.3$\\
\hline \hline
\end{tabular}
\end{center}
\end{table}

\begin{figure}[htbp]
\centering
\includegraphics[height=5cm]{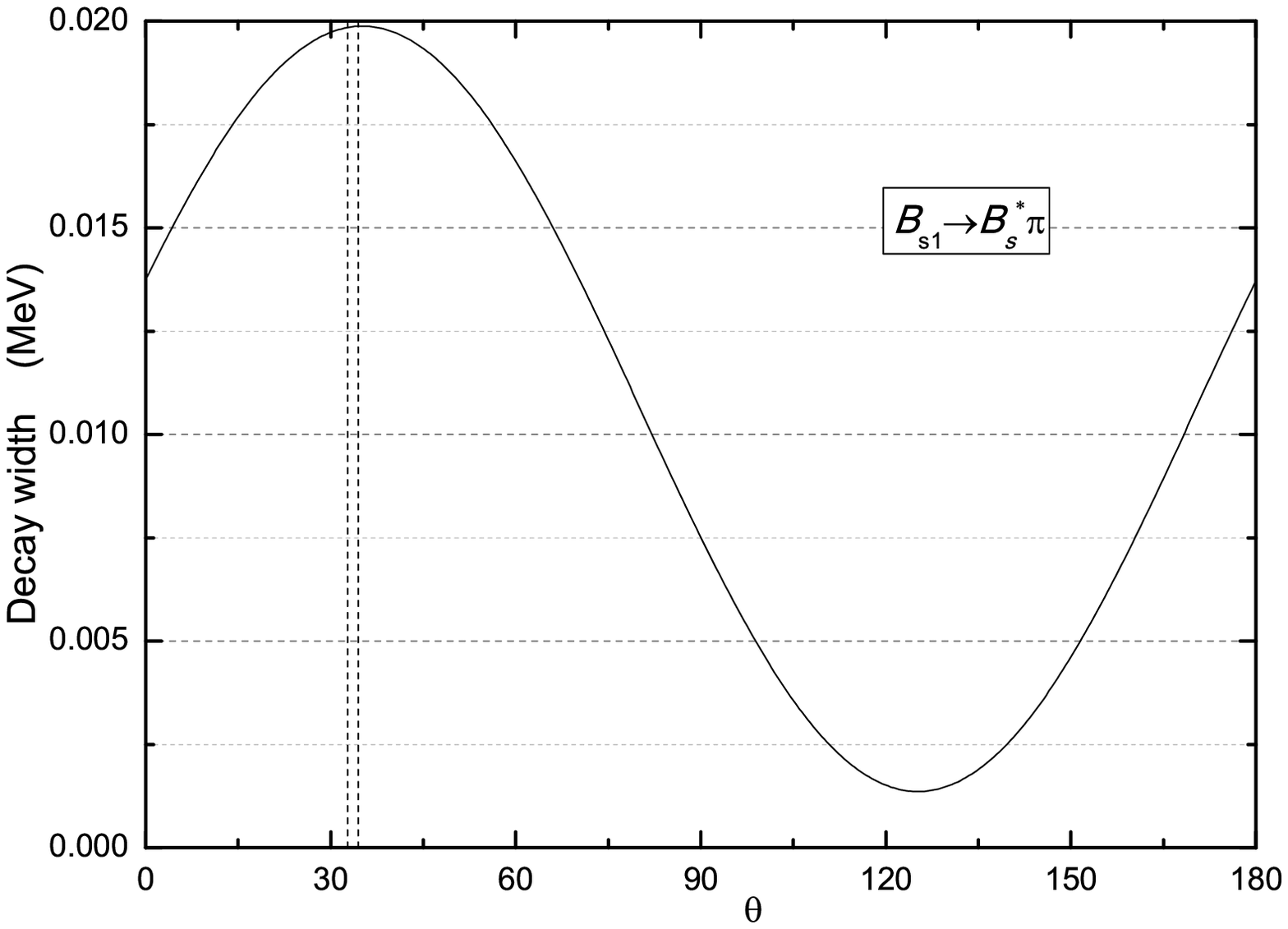}
\includegraphics[height=5cm]{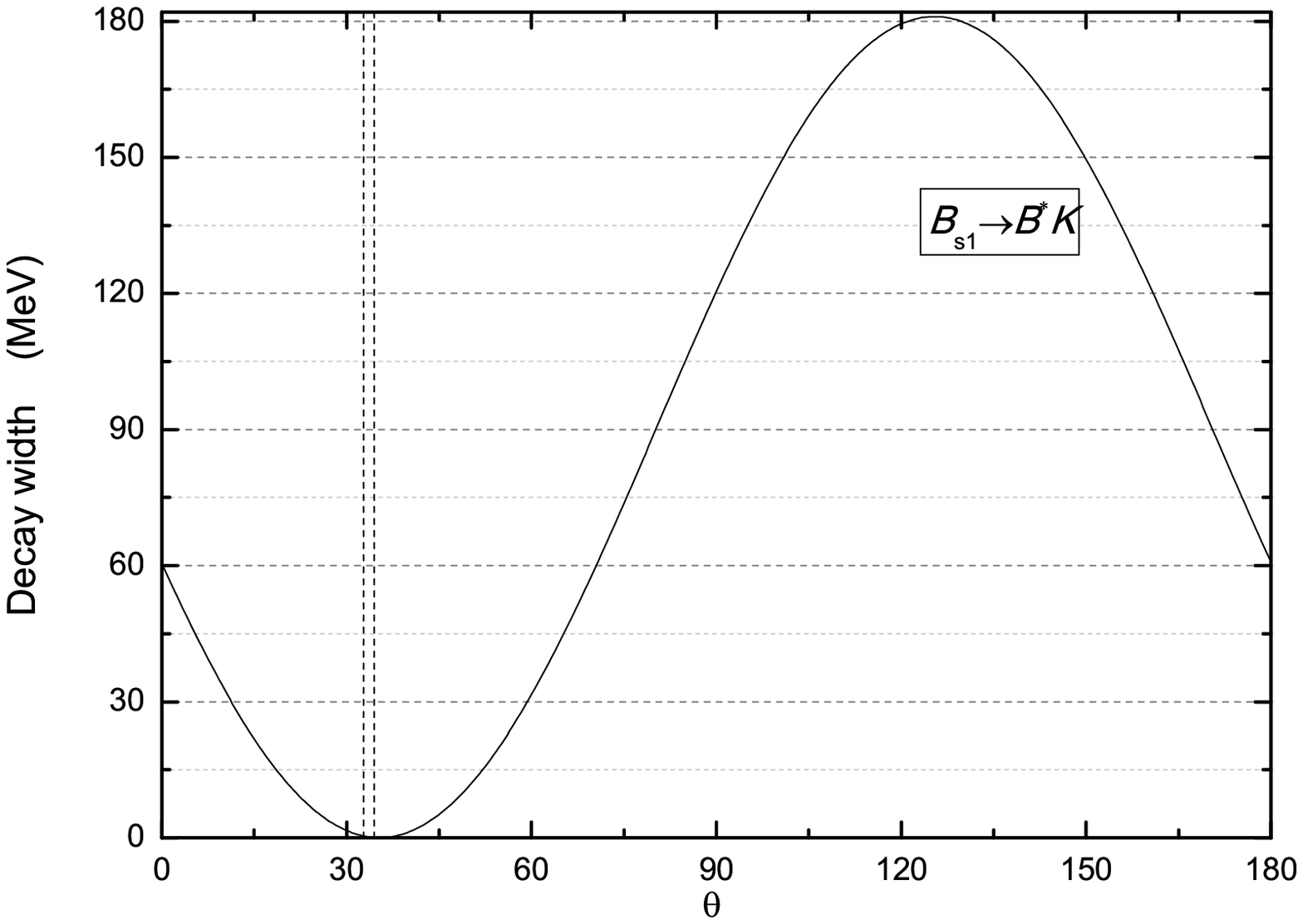}
\caption{\label{Bs1Bs11-Theta}The Decay widths of $B_{s1}^\prime$ and $B_{s1}$.}
\end{figure}

In order to compare the results in two conditions of mixing angle $\theta$ for $P$-wave $B^*_s$ mixing states $B_{s1}^\prime$ and $B_{s1}$,
we also show the corresponding strong decay widths of $B_{s1}^\prime$ and $B_{s1}$ in Table.~\ref{Bs1Bs11}.
For $B^\prime_{s1}$ meson, the OZI allowed strong decay is forbidden,
and the dominant strong decay of $B^\prime_{s1}$ meson is the isospin
symmetry violating decay via $\pi^0-\eta$ mixing~\cite{pieta}.
So the decay widths of $B^\prime_{s1}$ are very close for both of two conditions,
and the influence of the mixing angle is very small.
But for the decay width of $B_{s1}$ meson,
there is a big difference between two conditions,
the influence of the angle is very large.
In our calculation, we plot the decay widths of the
mixed states  $B^\prime_{s1}$ and $B_{s1}$ as functions of $\theta$ in Fig.~\ref{Bs1Bs11-Theta},
the results of two conditions are at the the bottom of curve and close to zero, which is the same as $D_{s1}^{\pm}$.
If we take the mixing angle $\theta=32.7^\circ$,
and get the decay width $\Gamma(B_{s1})=0.412$ MeV,
which is close to the experimental data with the large uncertainties.

In conclusion, we study the strong decays of $P-$wave mixing heavy-light $1^+$ states by the
improved B-S method in two conditions of mixing angle $\theta$: one is $\theta=35.3^\circ$;
another is considering the correction to mixing angle $\theta=35.3^\circ+\theta_1$.
We find that, for the $P$-wave $1^{\prime+}$ mesons ($D^{\prime0}_1$, $D^{\prime\pm}_1$, $D_{s1}^{\prime\pm}$, $B_1^{\prime0}$, $B^{\prime\pm}_1$ and $B^{\prime}_{s1}$)
and some $1^{+}$ states ($B_1^{0}$ and $B^{\pm}_1$),
the influence of mixing angle $\theta$ between $^3P_1$ and $^1P_1$ is very small, the results from two conditions are very close.
But for some of the $P$-wave $1^{+}$ mesons
($D^{0}_1$, $D^{\pm}_1$, $D_{s1}^{\pm}$ and $B_{s1}$),
influence of mixing angle $\theta$ between $^3P_1$ and $^1P_1$ is large,
especially for $D_{s1}^{\pm}$ and $B_{s1}$ mesons,
the discrepancy are very large between two conditions.
For $D_1^{0}$ and $D_1^{\pm}$ state, we take the mixing angle $\theta=26.7^\circ$ which is the best description of experimental values.
For $D_{s1}^{\pm}$ state, the result of the mixing angle $\theta=32.5^\circ$ is the best description of experimental value.
For $B_{s1}$ state, the result of the mixing angle $\theta=32.7^\circ$ is close to the experimental data with the large uncertainties.
In this paper, we have studied the strong decay of some special states which have not been discovered in experiment like:
$D^{\prime\pm}_1$, $B_1^{\prime0}$, $B^{\prime\pm}_1$ and $B^\prime_{s1}$,
that will provide the theoretical assistance to the future experiment.
We also investigate the strong decays of $D_{s1}^{\prime\pm}$, $B_1^{0}$, $B^{\pm}_1$ and $B_{s1}$ which have large uncertainties
for the experimental data, and give the predicted results,
there will need to be confirmed by the future experiments.

\noindent
{\Large \bf Acknowledgements}
This work was supported in part by
the National Natural Science Foundation of China (NSFC) under
Grant No.~11405004, No.~11405037, No.~11505039, No.~11575048 and the Key Scientific Research Projects in 2017 at  North Minzu University No.~2017KJ11.

\appendix{
\section{Instantaneous Bethe-Salpeter Equation}

In this section, we briefly review the Bethe-Salpeter equation and
its instantaneous one, the Salpeter equation.

The BS equation is read as~\cite{BS}:
\begin{equation}
(\not\!{p_{1}}-m_{1})\chi(q)(\not\!{p_{2}}+m_{2})=
i\int\frac{d^{4}k}{(2\pi)^{4}}V(P,k,q)\chi(k)\;, \label{eq1}
\end{equation}
where $\chi(q)$ is the BS wave function, $V(P,k,q)$ is the
interaction kernel between the quark and antiquark, and $p_{1},
p_{2}$ are the momentum of the quark 1 and anti-quark 2.

We divide the relative momentum $q$ into two parts,
$q_{\parallel}$ and $q_{\perp}$,
$$q^{\mu}=q^{\mu}_{\parallel}+q^{\mu}_{\perp}\;,$$
$$q^{\mu}_{\parallel}\equiv (P\cdot q/M^{2})P^{\mu}\;,\;\;\;
q^{\mu}_{\perp}\equiv q^{\mu}-q^{\mu}_{\parallel}\;.$$

In instantaneous approach, the kernel $V(P,k,q)$ takes the simple
form~\cite{Salp}:
$$V(P,k,q) \Rightarrow V(|\vec k-\vec q|)\;.$$

Let us introduce the notations $\varphi_{p}(q^{\mu}_{\perp})$ and
$\eta(q^{\mu}_{\perp})$ for three dimensional wave function as
follows:
$$
\varphi_{p}(q^{\mu}_{\perp})\equiv i\int
\frac{dq_{p}}{2\pi}\chi(q^{\mu}_{\parallel},q^{\mu}_{\perp})\;,
$$
\begin{equation}
\eta(q^{\mu}_{\perp})\equiv\int\frac{dk_{\perp}}{(2\pi)^{3}}
V(k_{\perp},q_{\perp})\varphi_{p}(k^{\mu}_{\perp})\;. \label{eq5}
\end{equation}
Then the BS equation can be rewritten as:
\begin{equation}
\chi(q_{\parallel},q_{\perp})=S_{1}(p_{1})\eta(q_{\perp})S_{2}(p_{2})\;.
\label{eq6}
\end{equation}
The propagators of the two constituents can be decomposed as:
\begin{equation}
S_{i}(p_{i})=\frac{\Lambda^{+}_{ip}(q_{\perp})}{J(i)q_{p}
+\alpha_{i}M-\omega_{i}+i\epsilon}+
\frac{\Lambda^{-}_{ip}(q_{\perp})}{J(i)q_{p}+\alpha_{i}M+\omega_{i}-i\epsilon}\;,
\label{eq7}
\end{equation}
with
\begin{equation}
\omega_{i}=\sqrt{m_{i}^{2}+q^{2}_{_T}}\;,\;\;\;
\Lambda^{\pm}_{ip}(q_{\perp})= \frac{1}{2\omega_{ip}}\left[
\frac{\not\!{P}}{M}\omega_{i}\pm
J(i)(m_{i}+{\not\!q}_{\perp})\right]\;, \label{eq8}
\end{equation}
where $i=1, 2$ for quark and anti-quark, respectively,
 and
$J(i)=(-1)^{i+1}$.

Introducing the notations $\varphi^{\pm\pm}_{p}(q_{\perp})$ as:
\begin{equation}
\varphi^{\pm\pm}_{p}(q_{\perp})\equiv
\Lambda^{\pm}_{1p}(q_{\perp})
\frac{\not\!{P}}{M}\varphi_{p}(q_{\perp}) \frac{\not\!{P}}{M}
\Lambda^{{\pm}}_{2p}(q_{\perp})\;. \label{eq10}
\end{equation}

With contour integration over $q_{p}$ on both sides of
Eq.~(\ref{eq6}), we obtain:
$$
\varphi_{p}(q_{\perp})=\frac{
\Lambda^{+}_{1p}(q_{\perp})\eta_{p}(q_{\perp})\Lambda^{+}_{2p}(q_{\perp})}
{(M-\omega_{1}-\omega_{2})}- \frac{
\Lambda^{-}_{1p}(q_{\perp})\eta_{p}(q_{\perp})\Lambda^{-}_{2p}(q_{\perp})}
{(M+\omega_{1}+\omega_{2})}\;,
$$
and the full Salpeter equation:
$$
(M-\omega_{1}-\omega_{2})\varphi^{++}_{p}(q_{\perp})=
\Lambda^{+}_{1p}(q_{\perp})\eta_{p}(q_{\perp})\Lambda^{+}_{2p}(q_{\perp})\;,
$$
$$(M+\omega_{1}+\omega_{2})\varphi^{--}_{p}(q_{\perp})=-
\Lambda^{-}_{1p}(q_{\perp})\eta_{p}(q_{\perp})\Lambda^{-}_{2p}(q_{\perp})\;,$$
\begin{equation}
\varphi^{+-}_{p}(q_{\perp})=\varphi^{-+}_{p}(q_{\perp})=0\;.
\label{eq11}
\end{equation}

For the different $J^{PC}$ (or $J^{P}$) states, we give the general form of
wave functions. Reducing the wave functions by the last
equation of Eq.~(\ref{eq11}), then solving the first and second equations in Eq.~(\ref{eq11}) to
get the wave functions and mass spectrum. We have discussed the
solution of the Salpeter equation in detail in Ref.~\cite{w1,mass1}.

The normalization condition for BS wave function is:
\begin{equation}
\int\frac{q_{_T}^2dq_{_T}}{2{\pi}^2}Tr\left[\overline\varphi^{++}
\frac{{/}\!\!\!
{P}}{M}\varphi^{++}\frac{{/}\!\!\!{P}}{M}-\overline\varphi^{--}
\frac{{/}\!\!\! {P}}{M}\varphi^{--}\frac{{/}\!\!\!
{P}}{M}\right]=2P_{0}\;. \label{eq12}
\end{equation}

 In our model, the instantaneous interaction kernel $V$ is Cornell
potential, which is the sum of a linear scalar interaction and a vector interaction:
\begin{equation}\label{vrww}
V(r)=V_s(r)+V_0+\gamma_{_0}\otimes\gamma^0 V_v(r)= \lambda
r+V_0-\gamma_{_0}\otimes\gamma^0\frac{4}{3}\frac{\alpha_s}{r}~,
\end{equation}
 where $\lambda$ is the string constant and $\alpha_s(\vec
q)$ is the running coupling constant. In order to fit the data of
heavy quarkonia, a constant $V_0$ is often added to confine
potential. To avoid the infrared divergence in the Coulomb-like one
and to correspond the fact that the confined linear interaction
should be also suppressed at large distance phenomenologically,
so it will be better to re-formulate the kernel as follows:
\begin{equation}
V_s(r)=\frac{\lambda}{\alpha}(1-e^{-\alpha r})~,
~~V_v(r)=-\frac{4}{3}\frac{\alpha_s}{r}e^{-\alpha r}~.
\end{equation}\label{vsvv}
 It is easy to
know that when $\alpha r\ll1$, the potential becomes to Eq.~(\ref{vrww}). In the momentum space and the C.M.S of the bound state,
the potential reads :
$$V(\vec q)=V_s(\vec q)
+\gamma_{_0}\otimes\gamma^0 V_v(\vec q)~,$$
\begin{equation}
V_s(\vec q)=-(\frac{\lambda}{\alpha}+V_0) \delta^3(\vec
q)+\frac{\lambda}{\pi^2} \frac{1}{{(\vec q}^2+{\alpha}^2)^2}~,
~~V_v(\vec q)=-\frac{2}{3{\pi}^2}\frac{\alpha_s( \vec q)}{{(\vec
q}^2+{\alpha}^2)}~,\label{eq16}
\end{equation}
where the running coupling constant $\alpha_s(\vec q)$ is :
$$\alpha_s(\vec q)=\frac{12\pi}{33-2N_f}\frac{1}
{\log (a+\frac{{\vec q}^2}{\Lambda^{2}_{QCD}})}~.$$ We introduce a small
parameter $a$ to
avoid the divergence in the denominator. The constants $\lambda$, $\alpha$, $V_0$ and
$\Lambda_{QCD}$ are the parameters that characterize the potential. $N_f=3$ for $\bar bq$ (and $\bar cq$) system.
}

\end{document}